\documentclass[a4paper,12pt]{article}
\usepackage{a4wide}
\usepackage{authblk}
\usepackage[dvips]{graphicx}
\usepackage{amsmath,amssymb}
\usepackage{bm}
\usepackage{subcaption}
\usepackage{siunitx}
\usepackage{url}
\usepackage{color}
\usepackage{cite}
\newcommand{\red}[1]{\textcolor{black}{#1}}

\newcommand{\tkmc}[1]{\textcolor{black}{#1}}

\title{A mode-in-state contribution factor based on Koopman operator and its application to power system analysis}
\author[1]{Kenji Takamichi\thanks{E-mail: sbb01101@edu.osakafu-u.ac.jp}}
\author[1]{Yoshihiko Susuki\thanks{Corresponding author. Present affiliation: Department of Electrical Engineering, Kyoto University, Katsura, Nishikyo-ku, Kyoto 615-8510 Japan. E-mail: susuki@ieee.org}}
\author[2]{Marcos Netto}
\author[1]{Atsushi Ishigame\thanks{Present affiliation: Department of Electrical and Electronic Systems Engineering, Osaka Metropolitan University, 1-1 Gakuen-cho, Naka-ku, Sakai, 599-8531 Japan}}
\affil[1]{Department of Electrical and Information Systems, Osaka Prefecture University, 1-1 Gakuen-cho, Naka-ku, Sakai, 599-8531 Japan}
\affil[2]{Power Systems Engineering Center, National Renewable Energy Laboratory, Golden, CO 80401 USA}
\date{}

\begin{document}
\maketitle
\begin{abstract}
This paper proposes a mode-in-state contribution factor for a class of  
nonlinear dynamical systems by utilizing spectral properties of the Koopman operator and sensitivity analysis. 
Using eigenfunctions of the Koopman operator for a target nonlinear system, we show that the relative contribution between modes and state variables can be quantified beyond a linear regime, where the nonlinearity of the system is taken into consideration.
The proposed contribution factor is applied to the numerical analysis of large-signal simulations for an interconnected AC/multi-terminal DC power system. 
\end{abstract}
\section{Introduction}
The so-called participation factor \cite{PF:ori} and contribution one \cite{CF} 
quantify mutual impacts between state 
variables and modes in multi-degree-of-freedom linear dynamical systems. 
Many groups of researches have studied definitions and interpretations of these factors from
viewpoints of systems theory and physics (see, e.g., \cite{PF:garofalo, PF:new}). 
Their generalizations to nonlinear systems are also reported in \cite{PF:nonlinear, PF:marcos}. 
Such generalizations are of fundamental concern in 
nonlinear systems theory and of technological significance in applications such as analysis and control of electric power systems \cite{PF:book}. 

In this paper, as a novel proposal, we introduce the mode-in-state contribution factor to a class of nonlinear dynamical systems. 
The mode-in-state contribution factor depends on (initial) states and is thus expected to extract global properties of nonlinear systems far from attractors.
Our theory is based on spectral theory of nonlinear dynamical systems, precisely speaking, spectral properties of the Koopman operators \cite{KMD:2005} and the idea of sensitivity analysis \cite{Kundur, PF:algebra}.
As introduced in Section~\ref{sec:KMD}, the Koopman operator is a linear infinite-dimensional operator defined for a nonlinear dynamical system and keeps full information on the nonlinear system: see, e.g., \cite{budisic12, ssk16, KMD:book}. 
The spectrum of the linear operator is the mathematical foundation of Koopman Mode Decomposition (KMD) \cite{KMD:2005, KMD:rowley}. 
Based on the KMD and the idea of sensitivity analysis, we introduce a data-driven mode-in-state contribution factor for a class of nonlinear systems. 
It should be noted that a KMD-based ``participation" factor has been proposed in \cite{PF:marcos}, which does not depend on states whereas our contribution factor does.
The introduced contribution factor is applied to the numerical analysis of large-signal simulations in an interconnected AC/Multi-Terminal DC (MTDC) power system \cite{Ohashi,KWMT2019,ISGT,KWMT2020}
in order to show its technological potential.
This paper is a substantially enhanced version of \cite{TKMC-NLP20,TKMC-NSW20}, former of which introduces a different definition of the mode-in-state contribution factor based on KMD.


\section{Proposal}

This section describes a new proposal of the mode-in-state contribution factor for nonlinear systems. 
For this, we introduce the Koopman operator and its spectrum leading to signal representation, called KMD (Koopman Mode Decomposition).

\subsection{Koopman operator and Koopman mode decomposition 
(KMD)
}
\label{sec:KMD}

Motivated by the application to power system analysis in this paper, we now consider continuous-time dynamical systems described by the following Differential-Algebraic Equation (DAE): 
\begin{equation}
\frac{d{\bm x}}{dt}={\dot{\bm x}} = {\bm F} ({\bm x} , {\bm y}), \qquad
{\bm 0} = {\bm G} ({\bm x},  {\bm y}),
\label{DAE}
\end{equation}%
where ${\bm x} \in {\mathbb R}^n$ is the vector of differential variables, 
${\bm y} \in {\mathbb R}^m$ is the vector of algebraic variables, and 
${\bm F}: {\mathbb R}^{n+m} \rightarrow {\mathbb R}^n$ and ${\bm G}: {\mathbb R}^{n+m} \rightarrow {\mathbb R}^m$ are 
given nonlinear vector-valued functions. 
We make the so-called regularity assumption \cite{SSK21} such that there exists a unique solution of DAE \eqref{DAE}, for which we address the dynamics in the set $\mathsf{R}$ defined as
\begin{equation}
    \mathsf{R}:=\{({\bm x},{\bm y})\in\mathbb{R}^n\times\mathbb{R}^m~|~
    {\bm G}({\bm x},{\bm y})={\bm 0},~
    \mathrm{det}({\rm D}_y{\bm G}({\bm x},{\bm y}))\neq 0\},
\end{equation}
where ${\rm D}_y{\bm G}$ is the Jacobian of $\bm G$ with respect to $\bm y$. 

It is shown in \cite{SSK21} that the Koopman operator is defined for DAE \eqref{DAE} with an asymptotically stable Equilibrium Point (EP). 
To see this, we denote by ${\bm S}_t({\bm x},{\bm y})$ the solution in $\mathsf{R}$ starting at $({\bm x},{\bm y})$ and converging to the stable EP (with a non-empty basin of attraction $\mathsf{A}\subset\mathsf{R}$) as $t\to\infty$. 
Then, the Koopman operator ${\bf U}_t$ is defined as a composition operator acting on a scalar-valued continuous function $g: {\mathsf A}
\rightarrow {\mathbb C}$, 
given by 
\begin{gather}
{\bf U}_t {g} := {g} \circ {\bm S_t}, \quad t\geq 0.
\end{gather}
The function $g$ is called \emph{observable}. 
Although DAE \eqref{DAE} and $\bm S_t$ are finite-dimensional and nonlinear, ${\bf U}_t$ is an infinite-dimensional but \emph{linear} operator acting on the space of functions.  
The linearity is utilized in the so-called Koopman operator framework for analysis and synthesis of nonlinear dynamical systems: see, e.g., \cite{budisic12,ssk16,KMD:book}.

One successful method in the framework is the KMD, in which signals generated by nonlinear systems are represented in terms of spectral properties of the Koopman operator. 
For this, the pair of \emph{Koopman eigenvalue} $\lambda\in\mathbb{C}$ and \emph{Koopman eigenfunction} $\phi_\lambda$ (a non-zero function defined on $\mathsf{A}$) is defined for $\mathbf{U}_t$ as follows:
\begin{equation}
{\bf U}_t \phi_\lambda = 
{\rm e}^{\lambda t} \phi_\lambda, \quad t\geq 0. 
\end{equation}
As proven in \cite{SSK21} under certain conditions, we expand a $K$-dimensional vector-valued observable ${\bm g}: \mathsf{A}\to\mathbb{R}^K$ in terms of Koopman eigenfunctions $\phi_j$, labeled by integer numbers $j$, and decompose the associated multivariate signal as follows:
\begin{equation}
{\bm g}(\bm S_t({\bm x},{\bm y})) 
= \sum_{j=1}^{\infty} {\rm e}^{\lambda_j t} \phi_j({\bm x},{\bm y}) {\bm V_j}, 
\qquad t\geq 0,
\quad ({\bm x},{\bm y})\in\mathsf{A},
\end{equation}
where $\lambda_j$ are the $j$-th Koopman eigenvalues, and ${\bm V_j}$ are called \emph{Koopman modes} for the expansion.
If we can take $\bm g(\bm x) = \bm x$, then the time evolution of the differential variables, denoted as $\bm x(t)=\bm S_t({\bm x},{\bm y})|_{\mathbb{R}^n}$, is decomposed as follows:
\begin{equation}
 \bm x(t) = \sum_{j=1}^{\infty} 
 {\rm e}^{\lambda_j t}\phi_j ({\bm x},{\bm y}) {\bm V_j}, 
 \quad t\geq 0.
 \label{KMD_x}
\end{equation}
Here, the $\bm y$ dependence in $\phi_j$ is resolved with the algebraic equation $\bm G(\bm x,\bm y)=\bm 0$. 
Precisely speaking, by introducing a transformation $\bm \varphi$ satisfying $\bm G(\bm x,\bm y=\bm \varphi(\bm x))=\bm 0$, in which its existence is guaranteed in $\mathsf{R}$ according to the implicit function theorem, it is possible to remove the $\bm y$ dependence as $\phi_j ({\bm x},{\bm \varphi}(\bm x))$ and to simply rewrite it as $\phi_j ({\bm x})$. 
For applications, it is numerically important to estimate the Koopman eigenvalues and eigenfunctions, and their computation is generally termed as the Dynamic Mode Decomposition (DMD) \cite{Kutz}. 
In particular, the so-called Extended DMD (EDMD) \cite{EDMD1
} is widely used for approximately deriving the Koopman eigenfunction as $\phi_j({\bm x}) \approx {\bm u}_j^\top {\bm \gamma}({\bm x})$, 
where $\bm u_j\in\mathbb{C}^m$ are related to left eigenvectors of an approximate 
matrix representation 
of the underlying Koopman operator, computed directly from time-series data of ${\bm x}(t)$. 
The $m$-dimensional vector-valued observable $\bm \gamma({\bm x})$ need to be designed 
for the computation.


\subsection{KMD-based mode-in-state contribution factor}

As a novel point of this paper, we introduce a mode-in-state contribution factor based on the KMD. 
This is based on the idea of sensitivity like \cite{Kundur, PF:algebra}.
Here, the term \emph{mode-in-state} indicates \emph{the contribution of modes in the time evolution of a single state, denoted by $x_k$, which are excited by a small change of initial value on the same state $x_k$}.  For this purpose, we consider an infinitesimal change of the state evolution $x_k(t)$, expressed as
\begin{equation}
  {\rm d}{x_k}(t) = \sum_{i=1}^n \frac{\partial x_k(t)}{\partial x_i}{\rm d}x_i,
  \label{CF1}
\end{equation}
where ${\rm d}x_i$ is a small change of the initial state $x_i$.
Now, to quantify the effect of the excitation of state $x_k$, we set ${\rm d}x_i = 0~(i \neq k)$. 
By substituting \eqref{KMD_x} into \eqref{CF1} with $\bm x=\bm x(0)$, \eqref{CF1} is formally re-written 
as follows:
\begin{equation}
  {\rm d}{x_k}(t) = \sum_{j=1}^\infty {\rm e}^{\lambda_j t} \frac{\partial \phi_j}{\partial x_k} ({\bm x}(0)) V_{j,k} {\rm d}x_k,
  \label{CF2}
\end{equation}
where $V_{j,k}$ is the $k$-th element of ${\bm V}_j$.
Here, from \eqref{CF2}, we define $\omega_{k,j}$ as the mode-in-state contribution factor between $j$-th mode and $k$-th state as follows:
\begin{equation}
  \omega_{k,j}:=\displaystyle \frac{\partial \phi_j}{\partial x_k} ({\bm x}(0))V_{j,k}.
  \label{CF3}
\end{equation}
This factor $\omega_{k,j}$ is dependent on the initial state ${\bm x}(0)$ and hence affected by the nonlinearity of the DAE system \eqref{DAE} through the Koopman eigenfunction $\phi_j({\bm x})$.
Numerically, using the estimated Koopman modes $\bm v_j$ 
and Koopman eigenfunctions ${\bm u_j}^\top {\bm \gamma}({\bm x})$, $\omega_{k,j}$ is approximately computed as follows:
\begin{equation}
  \left. \omega_{k,j} \approx v_{j,k} \sum_{\ell=1}^m u_{j,\ell} \frac{\partial \gamma_\ell}{\partial x_k}\right|_{{\bm x}={\bm x}(0)}.
   \label{CF4} 
\end{equation}

For the linear case, $\omega_{k,j}$ includes 
the mode-in-state participation factor in \cite{PF:ori}.  
%
Consider the linear DAE with $\bm F(\bm x, \bm y)={\bm A}_1{\bm x}+{\bm A}_2{\bm y}$ and $\bm G(\bm x, \bm y)={\bm A}_3{\bm x}+{\bm A}_4{\bm y}$ ( ${\bm A}_1\in\mathbb{R}^{n\times n}, {\bm A}_2\in\mathbb{R}^{n\times m}, {\bm A}_3\in\mathbb{R}^{m\times n}, {\bm A}_4\in\mathbb{R}^{m\times m}$).
It can be obtained by calculating the Jacobian matrices of the terms on the right-hand sides of the nonlinear DAE \eqref{DAE}.
If $\mathrm{det}(\mathrm{D}_y\bm G)=\mathrm{det}(\bm A_4)\neq 0$, then the dynamics of the differential variables are represented by the linear ordinary differential equation as
\begin{equation}
    \dot{\bm x}=(\bm A_1-\bm A_2\bm A^{-1}_4\bm A_3)\bm x
    =:\bm A\bm x.
    \label{eqn:linearODE}
\end{equation}
For this linear system, by assuming distinct $n$ eigenvalues $\lambda_j$ of $\bm A$ and associated left- (or right-) eigenvectors $\bm u_j$ (or $\bm v_j$), it is shown in \cite{Mezic13} that $\lambda_j$ and $\phi_j(\bm x)=\bm u^\top_j\bm x$ are principal eigenvalues and eigenfunctions of the Koopman operator. 
Since the Koopman mode $\bm V_j$ corresponds to $\bm v_j$ for the expansion of $\bm x$ in the linear system \cite{KMD:book}, 
\eqref{CF3} is written as
\begin{equation}
    \omega_{k,j}=u_{j,k}v_{j,k}, 
    \label{PF_linear}
\end{equation}
where it coincides with the participation factor \cite{PF:ori} derived for the linear system \eqref{eqn:linearODE}.

\section{Application to power system analysis}

\begin{figure*}[t]
 \begin{center}
  \includegraphics[width=1.0\textwidth]{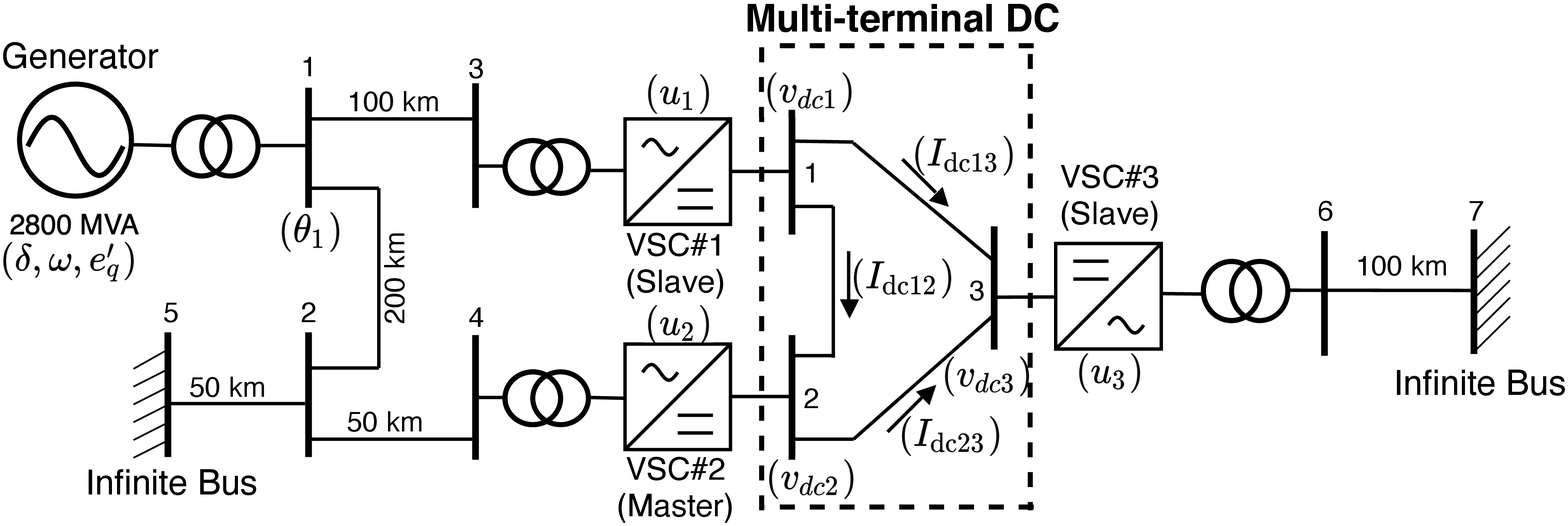}
  \caption{Single-line diagram of interconnected AC/multi-terminal DC power system in \cite{ISGT}
  }
  \label{ACMTDC}
 \end{center}
\end{figure*}

In this section, we apply the KMD-based mode-in-state contribution factor to the analysis of an interconnected AC/MTDC (Multi-Terminal DC) power system in \cite{Ohashi,KWMT2019,ISGT,KWMT2020}. 
The single-line  diagram of the AC/MTDC system is depicted in Fig.~\ref{ACMTDC}. 
As shown in \cite{KWMT2019,ISGT}, the mathematical model of the system is represented with DAE 
\eqref{DAE} with 12 differential variables $\bm x$ and 7 algebraic ones $\bm y$. 
The $\bm x$ includes the angular position $\delta$ of the synchronous generator, the deviation $\omega$ of rotor speed relative to a nominal angular frequency, the voltage $e'_q$ behind transient reactance in the generator, the DC current $I_{{\rm dc}ij}$ between DC bus $i$ and bus $j$, the DC voltage $v_{{\rm dc}i}$ at bus $i$, and the control inputs $u_i$ at Voltage Source Converter (VSC) $i$. 
Also, $\bm y$ includes the amplitudes and arguments (angles $\theta_i$) of voltage phasor at AC bus $i$. 
The variables focused in this paper are shown in Fig.~\ref{ACMTDC}, and the details of the model are described in \cite{Ohashi,KWMT2019,ISGT}.  
The model is intended for large-signal analysis of the AC/MTDC system, in which we need to evaluate the system dynamics beyond a neighborhood of a stable EP, that is to say, affected by the nonlinearity of the mathematical model.

The KMD-based mode-in-state contribution factor is now evaluated for numerical simulations of the model against disturbances in the generator voltage $e'_q$. 
The simulations are necessary for estimating the Koopman eigenvalues and eigenfunctions with 
EDMD. 
We generate $401$ initial conditions of $e'_q$ as $e'_q = e_q^{\prime \ast} + 0.0005\ell$ (for $\ell = 0,1,\ldots,400$), where $e_q^{\prime \ast}$ is the value at a stable EP, and we fix the initial conditions of the other differential variables in $\bm x$ at the stable EP. 
Then, for each initial condition, we simulate the time series of $\bm x$ with 300 
numbers of snapshots and sampling period 0.005\,s.  
Therefore, 120300 snapshots are used for the EDMD. 
In addition, based on \cite{Marcos2020}, 
the 22 
observables are selected as 
$\bm\gamma (\bm x)=[{\bm x}^\top, \cos(\delta-\theta_1), \sin(\delta-\theta_1), \cos2(\delta-\theta_1), \sin2(\delta-\theta_1), \omega \cos(\delta-\theta_1), \omega \sin(\delta-\theta_1), \omega \cos2(\delta-\theta_1), \omega \sin2(\delta-\theta_1), e_q^\prime \cos(\delta-\theta_1), e_q^\prime \sin(\delta-\theta_1)]^\top$.
The parameters are basically the same values as used in \cite{ISGT} although they do 
not show their complete list. 
In this paper there 
is no space here to show the complete set due to the limitation of space, but which can be provided upon request and will be shown in an archive.

Some of Koopman eigenvalues estimated by EDMD and eigevalues of the matrix $\bm A$ of \eqref{eqn:linearODE} derived via linearization of 
the mathematical model around a stable EP are presented in Table \ref{table:Eigenvalues}, which are related to the dynamics of the synchronous generator in the AC system. 
In Table \ref{table:Eigenvalues}, 
the estimated Koopman eigenvalues $\lambda_1$, $\lambda_{2,3}$ are close to the 
eigenvalues $\lambda_1^\prime$, $\lambda_{2,3}^\prime$ for the linearized model. 
This is valid in terms of the spectral characterization of the DAE with stable EP in \cite{SSK21}.
Furthermore, the Koopman eigenvalues $\lambda_{4,5}$ are likely the linear combination as $\lambda_1 + \lambda_{2,3} \times 2 = -3.06 \pm 27.2{\rm i}$.
The linear combination is known as the algebraic property of Koopman eigenvalues (see, e.g., \cite{SSK21}), hence $\lambda_{4,5}$ are generated by the nonlinearity of the model.

\begin{figure}[t] 
 \begin{minipage}[b]{0.32\linewidth}
    \centering
    \includegraphics[width=1.1\textwidth]{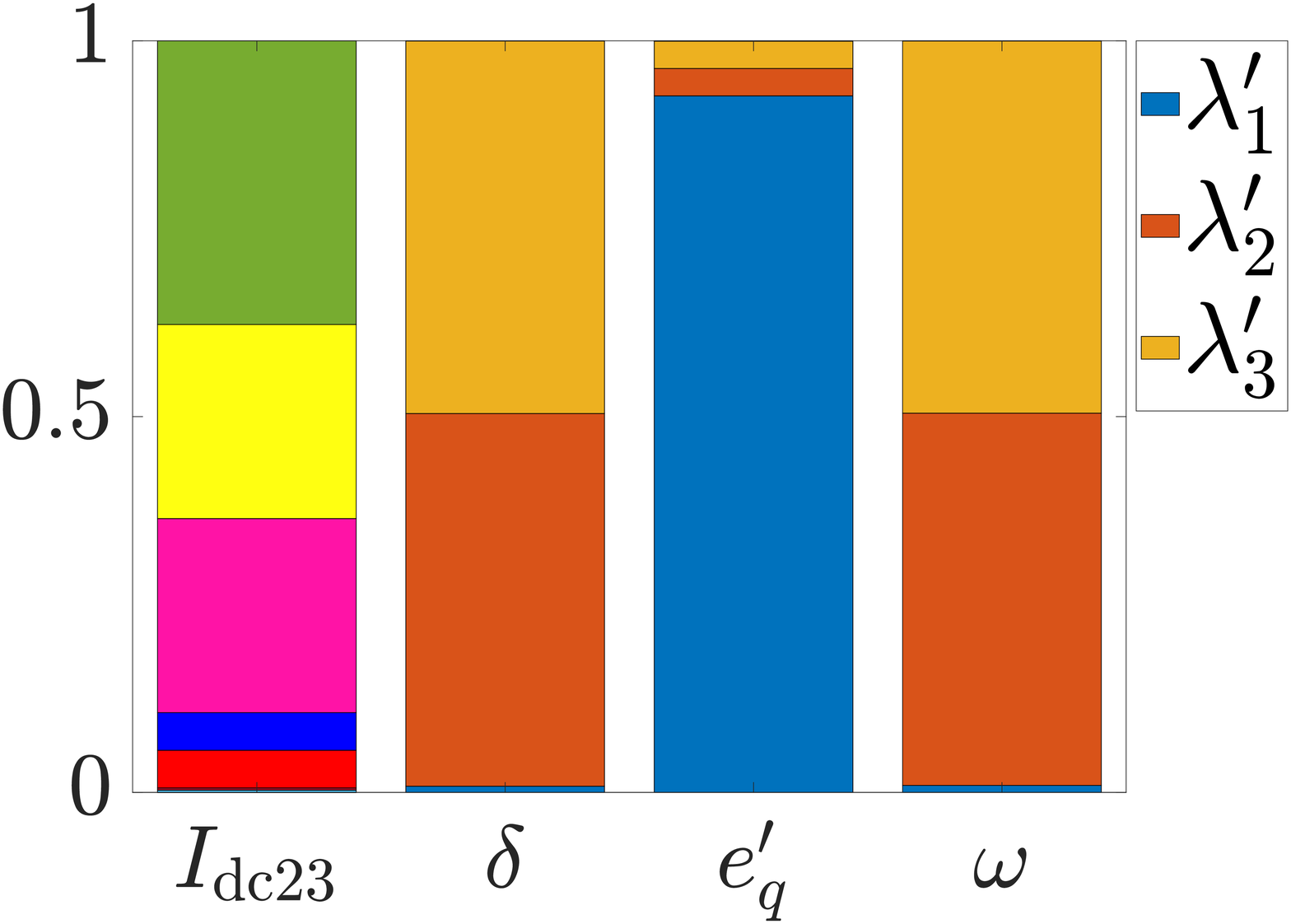}
    \subcaption{linearization}
    \label{PF}
  \end{minipage}
  \begin{minipage}[b]{0.32\linewidth}
    \centering
    \includegraphics[width=1.1\textwidth]{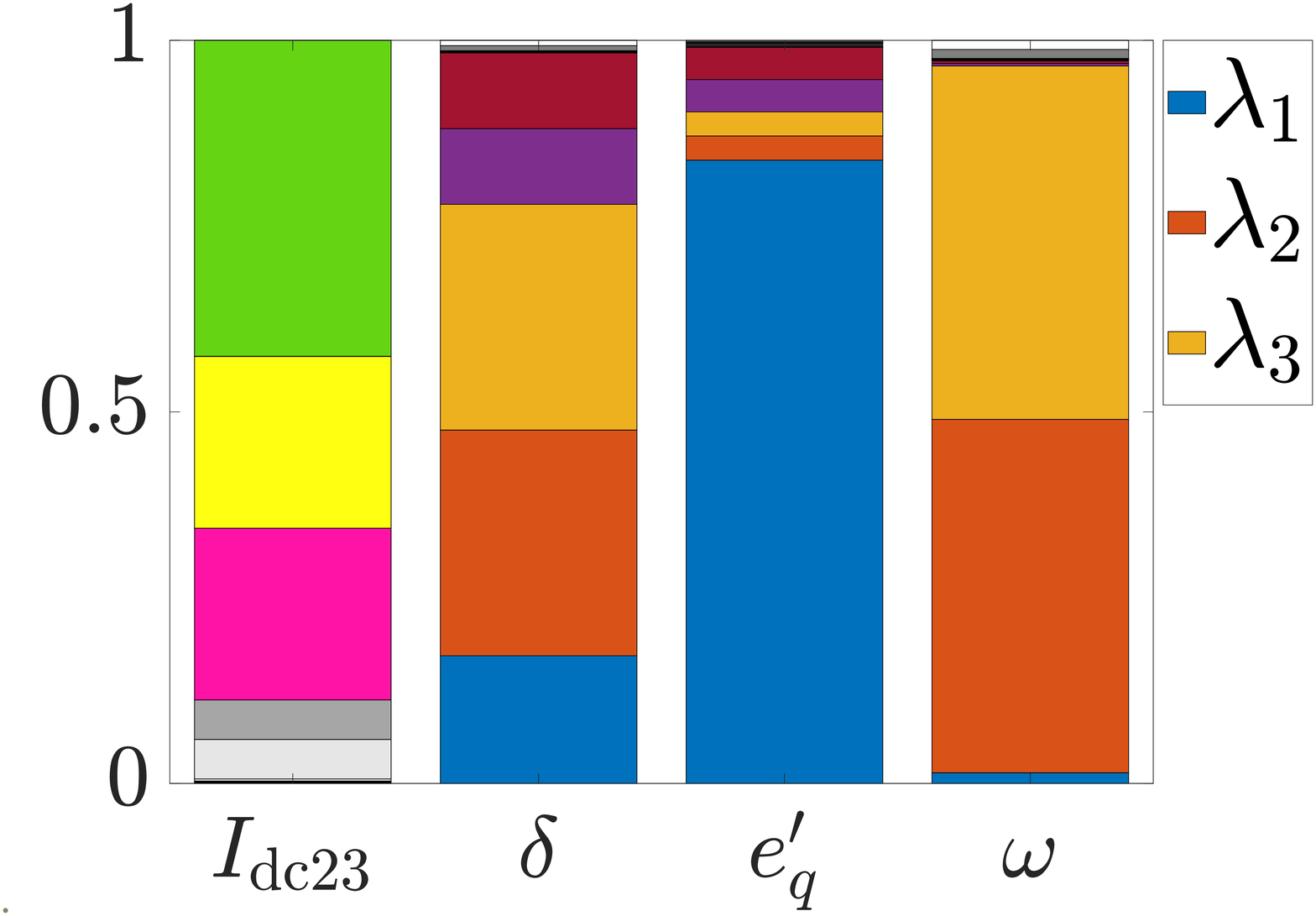}
    \subcaption{KMD-based} 
    \label{CF001}
  \end{minipage}
  \begin{minipage}[b]{0.32\linewidth}
    \centering
    \includegraphics[width=1.1\textwidth]{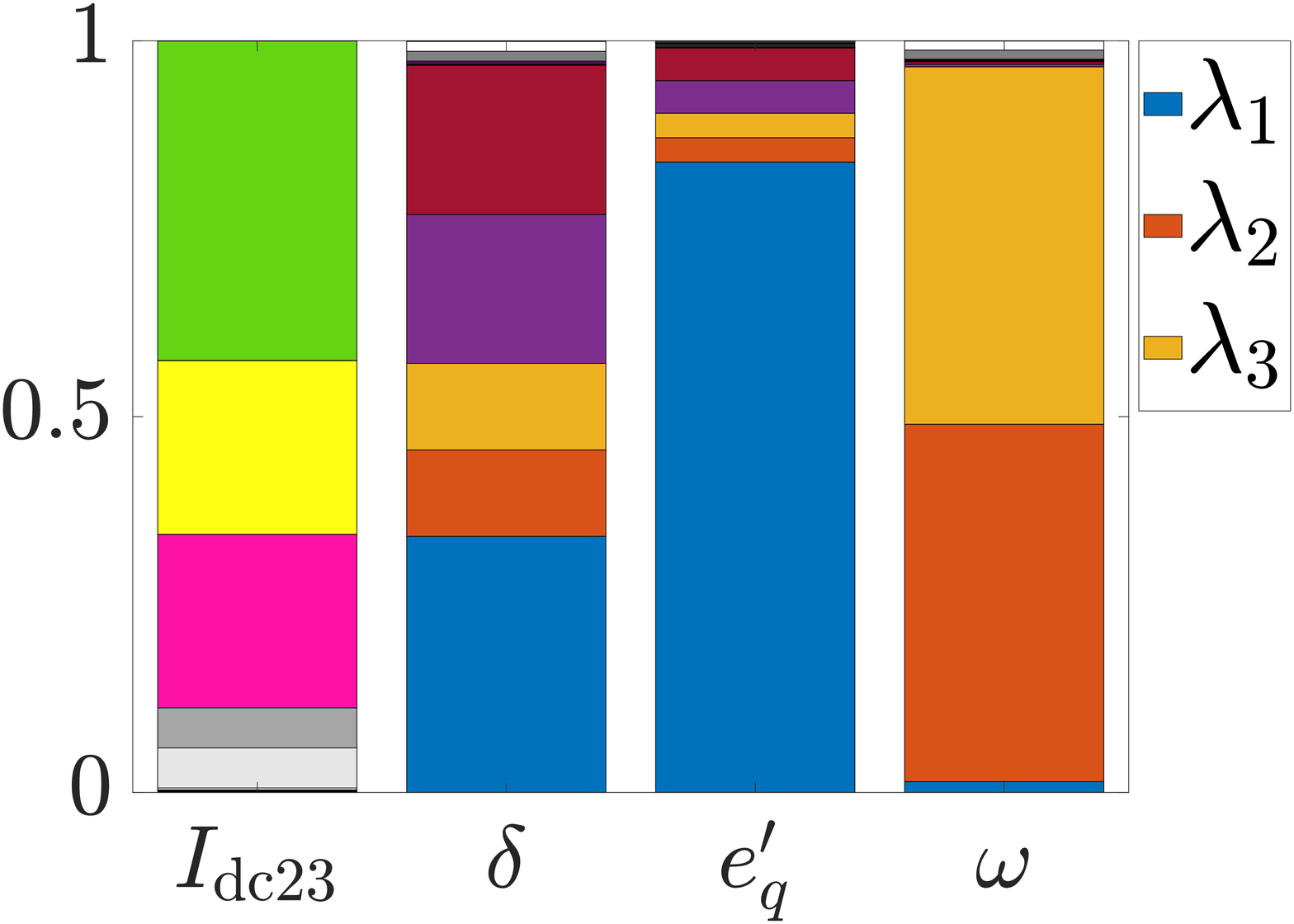}
    \subcaption{KMD-based} 
    \label{CF020}
  \end{minipage}
  \caption{
  Computation results on mode-in-state contribution and participation factors $\omega^\ast_{k,j}$ for the four variables. 
  \tkmc{The difference between Figs.~\ref{CF_ALL}(b) and (c) is that of the initial condition: $e'_q(0)=e^{\prime\ast}_q+0.01$ for (b) and $e'_q(0)=e^{\prime\ast}_q+0.2$ for (c). }
  Multiple colors are used to show the contributions of multiple modes (eigenvalues), where some of the eigenvalues in Table~\ref{table:Eigenvalues} are explicitly shown.}
  \label{CF_ALL}
\end{figure}

Finally, the KMD-based mode-in-state contribution factors are evaluated for different initial conditions. 
For clear comparison, motivated by \cite{Endegnanew}, we introduce the normalized magnitude of  contribution factor based on the absolute values of $\omega_{k,j}$ as follows:
\begin{equation}
   \omega_{k,j}^\ast := \frac{|\omega_{k,j}|}{\sum_{j=1}^{m}|\omega_{k,j}|}.
   \label{CF_normal} 
\end{equation}
Fig.~\ref{CF_ALL} shows the normalized magnitudes of contribution and participation factors for the four variables $I_{\rm dc23}, \delta, e'_q$, and $\omega$. 
The reason why the variables are chosen is that we emphasize the meaning and the relevance of the proposed factors as explained below.  
The mode-in-state participation factors with the conventional linear modal analysis calculated by \eqref{PF_linear} are presented in Fig.~\ref{CF_ALL}(a).
The KMD-based mode-in-state contribution factors are presented in Figs.~\ref{CF_ALL}(b) and (c). 
The difference between Figs.~\ref{CF_ALL}(b) and (c) is that of the initial condition: $e'_q(0)=e^{\prime\ast}_q+0.01$ for (b) and $e'_q(0)=e^{\prime\ast}_q+0.2$ for (c). 
In the figures, multiple colors are used to show the contributions of multiple modes, where some of the eigenvalues in Table~\ref{table:Eigenvalues} are explicitly shown. 
A clear difference in the contribution factors of $\delta$ among Figs.~\ref{CF_ALL}(a)-(c) is observed. 
In particular, the dependence on initial conditions in the figures (b) and (c) is caused by the nonlinearity of the model. 
In Fig.~\ref{CF_ALL}(b) with the small excitation, the linear mode $\lambda_{2,3}$ has a large contribution, whereas 
$\lambda_1$ has a small contribution. 
In Fig.~\ref{CF_ALL}(c) with the large excitation, $\lambda_{2,3}$ has lower a contribution and $\lambda_1$ has a larger contribution to $\delta$. 
Also, the contribution factors for $I_{\rm dc23}$ in the MTDC system do not change when the change of initial conditions in the AC system happens. 
This implies that disturbances that have strong impacts on the AC system do not affect the DC system remarkably. 
This computation is relevant in terms of control systems of VSC in the interconnected AC/MTDC system.
This is because the power system is controlled by the VSC in which dynamical interactions between the AC and MTDC systems are minimized. 

\begin{table}[t]
  \caption{
  Eigenvalues for linearized model and estimated Koopman eigenvalues, which are related to the dynamics of the synchronous generator in the AC system}
  \vspace*{2mm}
  \label{table:Eigenvalues}
  \centering
  \begin{tabular}{|c|c|c|c|}
    \hline
    \multicolumn{2}{|c|}{
    Eigenvalues for Linearized Model} & \multicolumn{2}{|c|}{Koopman Eigenvalues}\\
    \hline
    ${\lambda}_1^\prime$ & $-1.64$ 			& $\lambda_1$ & $-1.80$\\
    ${\lambda}_{2,3}^\prime$ & $-0.64$ $\pm$ 13.5i  & $\lambda_{2,3}$ & $-0.63$ $\pm$ $13.6$i \\
    					&					& $\lambda_{4,5}$ & $-2.98$ $\pm$ $26.2$i\\
    \hline
  \end{tabular}
\end{table}

\section{Conclusion}

A KMD-based mode-in-state contribution factor guided by the sensitivity analysis was proposed and applied to the analysis of an interconnected AC/MTDC power system.
Numerical results show that the proposed factor can capture the nonlinearity of the model and the physical property of the target power system. 
Several follow-up studies on this paper are possible.
A KMD-based state-in-mode contribution factor should be pursued to quantify the mutual impact between one mode and each state. 
Also, an application of the proposed factor to a multi-machine power system exhibiting an inter-area oscillation is interesting and important in practical viewpoints.

\section*{Acknowledgments}
The work was partially supported by JST-PRESTO Grant Number JPMJPR1926 and JST-SICORP Grant Number JPMJSC17C2. 


\section*{Appendix \red{to the Archival Manuscript}}
We show the 
parameter values \red{for numerical simulations} of the AC/multi-terminal DC system used in Section 3.
The same parameters and mathematical model are used as in the simulation\red{s} in \cite{ISGT} although they do not show their complete parameter list.
The parameters on the AC and DC grids are mainly based on \cite{Ohashi}, and the parameters on the \red{synchronous} generator are mainly based on \cite{KWMT2019}.
Table \ref{table:Parameters} shows the descriptions and values of the parameters on the AC grid, DC grid, and generator.
\red{For per-unit system,} 
we \red{use the following base quantities}: 
Base power of the generator is 2800\,MVA; 
Base power \red{and voltage} of the AC \red{grid} 
is 1000\,MVA and 
is 500\,kV; 
Base power \red{and voltage} of the DC \red{grid} 
is 1000\,MVA and 
500\,kV; and 
Base angular frequency of \red{both} the AC and DC \red{grids} 
is $50 \times 2\pi$ radian \red{per second}.
\newpage

\begin{table}[t] 
  \caption{
  Descriptions and values of the parameters on \red{the} AC grid, DC grid, and 
  generator.
  The parameters are mainly based on \cite{Ohashi, KWMT2019, ISGT}. 
  }
  \vspace*{2mm}
  \label{table:Parameters}
  \centering
  \begin{tabular}{ccc}
  \hline
    \hline
   Parameter & Description & Value\\
    \hline
    $r_{\rm AC}$ & Resistance of AC \red{line} 
    per length [\si{\ohm}/km]	& 0.0261\\ \hline
    $l_{\rm AC}$ & Reactance of AC \red{line} 
    per length [mH/km]	& 0.83\\ \hline
    $r_{\rm DC}$ & Resistance of DC \red{line} 
    per length [\si{\ohm}/km]	& 0.0192\\ \hline
    $l_{\rm DC}$ & Reactance of DC \red{line} 
    per length [mH/km]	& 0.24\\ \hline
    $c_{\rm DC}$ & Capacitance of DC \red{line} 
    per length [$\mu$F/km]	& 0.152\\ \hline
    $c_{\rm DC conv}$ & Smoothing capacitance \red{for VSC} [$\mu$F/km]	& 75\\ \hline
    $K_V$ & \red{Constant for voltage conversion} 
    & 2\\ \hline
    $\sqrt{3}K_I$ & \red{Constant for current conversion} 
    & 1\\ \hline
    $G$ & \red{Gain constant of controller for VSC} 
    & --1\\ \hline
    $T$ & Time constant \red{of controller for VSC} 
    & 0.001\\ \hline
    $P_{\rm DC(ref)1}$ & Reference power at VSC1 & 0.2\\ \hline
    $V_{\rm DC(ref)2}$ & Reference voltage at VSC2 & 1.0\\ \hline
    $P_{\rm DC(ref)3}$ & Reference power at VSC3 & --0.3\\ \hline
    $X_{d}$ & 
    $d$-axis synchronous reactance \red{in generator} 
    & 1.79\\ \hline
    $X_{q}$ & 
    $q$-axis synchronous reactance \red{in generator} 
    & 1.77\\ \hline
    $X^{\prime}_{d}$ & 
    $d$-axis transient reactance \red{in generator} 
    & 0.3\\ \hline
    $E_{fd}$ & Constant voltage behind $d$-axis synchronous reactance & 1.70\\ \hline
    $P_{m}$ & 
    Mechanical power injection \red{to generator} & 0.5\\ \hline
    $D$ & 
    Damping coefficient \red{in generator} & 1.0\\ \hline
    $H$ & Inertia constant of rotor \red{in generator} & $0.89 \times 100\pi$\\ \hline
    $T^{\prime}_{d0}$ & $d$-axis transient open-circuit time constant & $1.2 \times 100\pi$\\ \hline
    \hline
  \end{tabular}
\end{table}





\begin{thebibliography}{99}
\bibitem{PF:ori} I. J. P\'{e}rez-Arriaga et al., 
{\it IEEE Trans. Power App. Syst}., vol. PAS-101, no. 9, pp. 3117--3125, Sep. 1982.
DOI:10.1109/TPAS.1982.317524
%
\bibitem{CF} S. K. Starrett et al., 
{\it Proc. Int. Symp. Nonlin. Theory Appl}., vol. 2, p. 523--538, Dec. 1993.
%
%
\bibitem{PF:garofalo} F. Garofalo et al., 
{\it IFAC Proc. Volumes}, vol. 35, no. 1, pp. 125--130, 2002. 
DOI:10.3182/20020721-6-ES-1901.00182

\bibitem{PF:new} W. A. Hashlamoun et al., 
{\it IEEE Trans. Automat. Contr}., vol. 54, no. 7, pp. 1439--1449, July 2009, DOI: 10.1109/TAC.2009.2019796

\bibitem{PF:nonlinear} B. Hamzi and E. H. Abed, 
{\it Proc. 53rd IEEE Conference on Decision and Control}, pp. 43--48, 2014. 
DOI:10.1109/CDC.2014.7039357

\bibitem{PF:marcos} M. Netto et al., 
{\it IEEE Contr. Syst. Lett}., vol. 3, no. 1, pp. 198--203, Jan. 2019. 
DOI:10.1109/LCSYS.2018.2871887

\bibitem{PF:book} J. H. Chow, {\it Power System Coherency and Model Reduction}, Springer, 2013.

\bibitem{KMD:2005}I. Mezi\'{c}, 
{\it Nonlinear Dyn}., vol. 41, pp. 309--325, 2005. 
DOI:10.1007/s11071-005-2824-x

\bibitem{Kundur} P. Kundur, \emph{Power System Stability and Control}, McGraw-Hill, 1994.

\bibitem{PF:algebra} G. Tzounas et al., 
{\it IEEE Trans. Power Syst}., vol. 35, no. 1, pp. 742--750, 2020. 
DOI:10.1109/TPWRS.2019.2931965

\bibitem{budisic12} M. Budi\v{s}i\'{c} et al., {\it CHAOS}, vol. 22, no. 4, article \#047510, December 2012. DOI:10.1063/1.4772195

\bibitem{ssk16} Y. Susuki et al., 
{\it NOLTA, IEICE}, vol. 7, Issue 4, pp. 430--459, 2016. 
DOI:10.1587/nolta.7.430

\bibitem{KMD:book} A. Mauroy et al., 
{\it The Koopman Operator in Systems and Control: Concepts, Methodologies, and Applications}, Springer, 2020.
\bibitem{KMD:rowley} C. W. Rowley et al., {\it J. Fluid Mech}., vol. 641, pp. 115--127, 2009.  
DOI:10.1017/S0022112009992059

\bibitem{Ohashi} Y. Ohashi et al., 
{\it IEE-Japan Tech., Rep}., \#PSE-19-002, 2019 (in Japanese).

\bibitem{KWMT2019} N. Kawamoto et al., 
{\it NOLTA, IEICE}, vol. 11, no. 4, pp. 610--623, 2020. 
DOI:10.1587/nolta.11.610

\bibitem{ISGT} Y. Susuki et al., 
{\it Proc. 2020 IEEE PES Innovative Smart Grid Technologies Europe (ISGT-Europe)}, pp. 945--949, 2020. 
DOI:10.1109/ISGT-Europe47291.2020.9248890


\bibitem{KWMT2020} N. Kawamoto et al., 
{\it NOLTA, IEICE}, vol. 12, no. 4, pp. 711--717, 2021. 
DOI:10.1587/nolta.12.711 

\bibitem{TKMC-NLP20} K. Takamichi et al., 
{\it IEICE Tech. Rep}., vol. 121, no. 61, NLP2021-8, pp. 34--39, June 2021 (in Japanese).

\bibitem{TKMC-NSW20} K. Takamichi et al., 
{\it Proc. 2021 Nonlinear Science Workshop}, December 2021. 

\bibitem{SSK21} Y. Susuki, 
{\it Preprints of Third IFAC Conference on Modeling, Identification and Control of Nonlinear Systems (IFAC MICNON 2021)}, pp. 361--365, 2021.

\bibitem{Kutz} J. Kutz et al., 
{\it Dynamic Mode Decomposition: Data-Driven Modeling of Complex Systems}, SIAM, 2016. 

\bibitem{EDMD1}M. O. Williams et al., 
{\it J. Nonlinear Sci}., vol. 25, no. 6, pp. 1307--1346, 2015. 
DOI:10.1007/s00332-015-9258-5 


\bibitem{Mezic13} I. Mezi\'{c}, 
{\it Ann. Rev. Fluid Mech}., vol. 45, no. 1, pp. 357--378, 2013. 
DOI:10.1146/annurev-fluid-011212-140652

\bibitem{Marcos2020} M. Netto et al., {\it IEEE Contr. Syst. Lett}., vol. 5, no. 6, pp. 1868--1873, December 2021. DOI:10.1109/LCSYS.2020.3047586

\bibitem{Endegnanew} A. G. Endegnanew et al., 
{\it Proc. 2017 Twelfth International Conference on Ecological Vehicles and Renewable Energies (EVER)}, pp. 1--8 (2017). DOI:10.1109/EVER.2017.7935937
\end{thebibliography}
\end{document}